\documentclass[11pt,twoside]{article}

\usepackage{asp2006}
\usepackage{epsf}
\usepackage{psfig}
\usepackage{lscape}

\begin{document}
\title{The History of Star forming Galaxies and their environment as seen by {\it Spitzer}: A Review}   
\author{Isaac Roseboom$^{1}$, Seb Oliver$^{1}$, Duncan Farrah$^{1}$, Mark Frost$^{1}$}   
\affil{$^{1}$Department of Physics and Astronomy, University of Sussex, Falmer, BN1 9QH, United Kingdom}    

\begin{abstract} 
The advent of the {\it Spitzer} Space Telescope has revolutionized our understanding of the history of star formation and galaxy mass assembly in the Universe. The tremendous leap in sensitivity from previous mid-to-far IR missions has allowed {\it Spitzer} to perform deeper, and wider, surveys than previously possible at these wavelengths. In this review I highlight some of the key results to come out of these surveys, and the implications these have for current models of galaxy formation and evolution.

\end{abstract}


\section{Introduction}
{\it Spitzer}, with its complementary suite of near-to-far infrared instruments, has made important contributions to virtually every aspect of astrophysics in the 6 years since its launch. To satisfactorily review all of its accomplishments, even in the single field of galaxy formation and evolution, would be a tremendous task. Thus this all-too-brief review will cover only selected highlights of results coming from some of the key extragalactic legacy surveys, in particular; the Great Observatories Origins Deep Survey (GOODS; Dickinson et al. in prep), the Spitzer Wide-area InfraRed Extragalactic survey (SWIRE; Lonsdale et al. 2003), The Far Infrared Deep Extragalactic Legacy survey (FIDEL; PI Dickinson) and {\it Spitzer} observations taken as part of the Cosmic Evolution Survey (COSMOS; Scoville et al. 2007). In particular a focus on statistical measures, such as the luminosity function or two-point correlation function, is made as these are areas in which {\it Spitzer}'s ability to survey large areas to exceptional depth has made the biggest impact. 

This review is broken down into three distinct parts; 1. the star formation history of galaxies, 2. Properties of star forming galaxies, and 3. The environment of star forming galaxies, as seen by {\it Spitzer}.
\section{Star formation history of Galaxies}
\subsection{Cosmic Infrared Background}
The most direct, albeit crude, measure of the mid-to-far IR properties of galaxies is to simply measure the total emission across the sky, which forms the Cosmic Infra-red Background (CIB; Hauser et al. 1998). While some of this emission comes from AGN, the bulk of this can be directly attributed to star forming processes in distant galaxies. {\it Spitzer} has enabled, for the first time, a large fraction of the galaxies which make up the CIB at different wavelengths to be resolved. At 24$\mu$m Papovich et al. (2004) directly observed the turnover in the number counts at $\sim 0.2$mJy and hence directly estimate a lower limit to the CIB at 24$\mu$m to be $\nu I_{\nu}=2.7^{+1.1}_{-0.7}$ nW m$^{-2}$ sr$^{-1}$, consistent with {\it COBE} DIRBE upper limit of  $\nu I_{\nu}<7$ nW m$^{-2}$ sr$^{-1}$ (Kashlinsky \& Odenwald 2000; Hauser \& Dwek 2001). Of this background $\nu I_{\nu}=1.9\pm0.6$ nW m$^{-2}$ sr$^{-1}$ ($\sim$70\%) is contributed by sources $>60\mu$Jy which are directly resolved by {\it Spitzer} MIPS.

However even with the improved sensitivity of {\it Spitzer} a large fraction of the 70 \& 160 $\mu$m background cannot be resolved directly into individual sources. Dole et al. (2004) show that only $\sim$20\% ($\sim$10\%) of the CIB at 70$\mu$m (160$\mu$m) is contributed by sources resolved by MIPS surveys at those wavelengths. One way to get around this is to ``stack'' the images at 70 (\& 160) $\mu$m based on the known positions in shorter wavelength data. Dole et al (2006) performed stacking of the MIPS 70 \& 160 $\mu$m images in three fields (CDFS, HDFN \& Lockman Hole) based on the known positions of 24$\mu$m sources. The 24$\mu$m sources were distributed into 20 bins of descending flux density. Subsequently the 70 \& 160 $\mu$m micron MIPS images were co-added at the positions of the 24$\mu$m sources in each flux density bin. In this way the contribution to the 70$\mu$m and 160$\mu$m background as a function of 24$\mu$m flux can be assessed. Again using the observed behaviour of the 24$\mu$m number counts a robust lower limit on the 70$\mu$m \& 160$\mu$m background of 7.1$\pm1.0$ \& 13.4$\pm1.7$  nW m$^{-2}$ sr$^{-1}$. Sources resolved at 24$\mu$m (i.e. $>60\mu$Jy) contribute $\sim80$\% of this background at 70 \& 160 $\mu$m. 
\subsection{Number counts \& Luminosity function of star forming galaxies}
One of the greatest surprises to come out of the earliest {\it Spitzer} observations was the great discrepancies between the observed number counts at MIPS wavelengths and those predicted by pre-{\it Spitzer} IR population models. In particular Papovich et al. (2004) found that the peak in the 24$\mu$m counts was at a much lower flux density than predicted from phenomenological constrained to match ISO 15$\mu$m surveys. Papovich et al., and subsequent studies, attribute this discrepency to a previously unobserved population of $z=$1--3 Luminous Infra-Red Galaxies (LIRGs) with IR luminosities in the range $10^{11}L_{\odot}<L_{IR}<10^{12}L_{\odot}$ (Shupe et al. (2008); Pearson et al. in prep).

At longer wavelengths the sensitivity of {\it Spitzer} is not great enough to probe the high redshift 70$\mu$m \& 160$\mu$m populations and hence the number counts at these wavelengths are generally much closer to those predicted by pre-{\it Spitzer} models (see Lagache et al. 2004; Frayer et al. 2006; Frayer et al. 2009). 

Finally Le Borgne et al (2009) have used a combination of the published number counts at all {\it Spitzer} MIPS wavelengths, in combination with the published 15$\mu$m counts from ISO and the 850$\mu$m counts from SCUBA to find the best-fit star formation history of the Universe. Their approach involves a non-parametric inversion of the counts at the 5 wavelengths. Specifically they blindly construct bolometric IR luminosity functions which can match the observed number counts, assuming the flux at each wavelengths is given by an phenomenological set of far-IR SEDs. The range of IR luminosity functions and redshifts which are allowed by the data are then collapsed to give the range of plausible star formation rate densities (SFRD) of the Universe. They find good agreement to direct measures of the SFRD as compiled by Hopkins \& Beacom (2006), with the conclusion that LIRGS at $z=$1--2 and Ultra Luminous Infra-Red Galaxies (ULIRGS: $10^{12}L_{\odot}<L_{IR}<10^{13}L_{\odot}$) at $z\sim2$ are the dominant contributors to the SFRD, and hence the CIB.

While number counts are a useful tool for constraining population models, the degeneracy between luminosity and redshift can limit their interpretability. Fortunately great effort has been put into large spectroscopic campaigns, as well as the production of accurate photometric redshifts, in the deepest {\it Spitzer} survey fields.

Taking advantage of data from the VVDS,GOODS,COMBO-17 and deep 24$\mu$m surveys in the CDFS field, Le Floch et al. (2005) were amongst the first to show that the space density of LIRGs increases by a factor of $\sim100$ from $z=$0--1. This result has been subsequently confirmed by Magnelli et al. (2009) using both MIPS 24$\mu$m and 70$\mu$m data taken as part of the FIDEL survey. At higher redshift ($z>2$) Caputi et al. (2007) find that an even greater increase ($>100$ times) in the space density of ULIRGs from $z=$0--2. These results match up well with the picture painted by the analysis of the number counts, with LIRGs dominating the far-IR luminosity density, and hence star formation density, at $z\sim1$ and ULIRGs at $z>2$. However it is worth noting that most of this work relies strongly on the 24$\mu$m data from {\it Spitzer} which is probing different parts of the mid-IR SED at different redshifts. This is particularly a problem at $z>2$ where the 24$\mu$m band is probing $\sim8\mu$m rest-frame, a region which can often be contaminated by PAH and molecular line emission.

\section{Properties of star forming galaxies}
While {\it Spitzer} MIPS data has been key to uncovering the evolution in star formation since $z\sim2$, by combination with overlapping {\it Spitzer} IRAC and ground-based optical/near-IR data the nature of galaxies hosting this star formation can be determined.

Using the overlap between the SWIRE survey and the Sloan Digital Sky Survey (SDSS), Davoodi et al. (2006) showed the disconnect between the optical and mid-far infrared properties of local galaxies ($z<0.2$). Specifically they found a significant fraction ($\sim 30$\%) of red sequence galaxies show an excess in their mid infrared emission, defined as $\log(L_{24}/L_{3.6})>-0.7$). Roughly half of this red, mid-IR strong, population are determined to be star-forming systems from emission line diagnositics (Baldwin et al. 1981), with the other half coming from AGN processes. 

In addition to optical colour (essentially a proxy for stellar age), determination of the stellar mass of star forming systems, and its evolution with redshift, is of particular interest.

Zheng et al. (2007) used a combination of COMBO-17 data with {\it Spitzer} IRAC \& MIPS observations to determine the evolution of the specific star formation rate (SSFR: SFR/M$_{\odot}$) from $z=$0--1. They find clear evidence of a significant decrease in the SSFR across a wide range of stellar mass (10$^{9}$M$_{\odot}$--10$^{11}$M$_{\odot}$) from $z=1$ to present. While the SSFR is found to decrease with increasing mass, as postulated by ``downsizing'' the evolution of the SSFR is similar across the range of stellar mass. Thus it is clear that the mechanism driving this evolution is a) different from that at high-$z$, otherwise massive galaxies would never form, and b) the same for all galaxies at $z<1$. Zheng et al. suggest that the decline in available infalling cool gas, as opposed to environmental or feedback effects, is the underlying cause of this behaviour.

Damen et al. (2009) and Perez-Gonzalez et al. (2008) have utilised the deeper FIDEL MIPS data in CDFS \& the extended Groth strip respectively to investigate the evolution of the SSFR out to $z\sim2$. Damen et al. find good agreement with the Zheng et al. results at low-$z$, again seeing a strong increase in the SSFR with redshift across all stelar masses. With the greater depth of the FIDEL data they are able to track the SSFR out to $z=2$ for the most massive galaxies ($>10^{11}$M$_{\odot}$). The SSFR in massive galaxies is observed to increase by a factor of nearly 10 from $z=$1--2, qualitatively in line with the downsizing paradigm, although again the slope of the evolution in the SSFR is found to be independent of stellar mass. Perez-Gonzalez et al. find a similar increase in the SSFR with redshift, with an observed increase of nearly 100$\times$ from $z=$0--2. Interestingly they also investigate the morphological dependence of this evolution. Seperating spheriods from disk dominated systems via use of the Sersic index differential evolution in the SSFR is observed, with the spheriod dominated systems experiencing a much stronger increase in SSFR with redshift. 

\begin{figure}[!ht]
   \resizebox{1.2\hsize}{!}{
     \includegraphics*{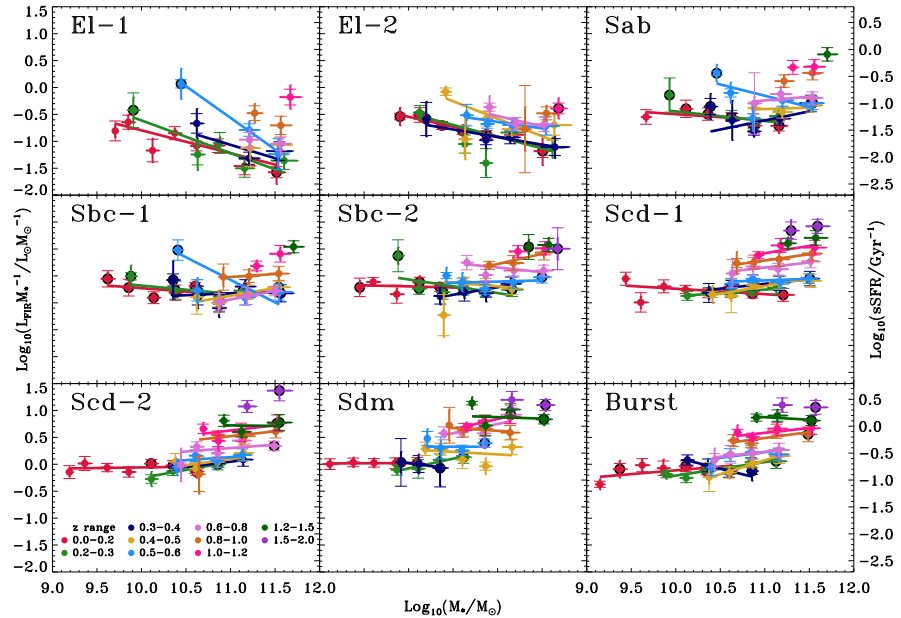}
   }
\caption{SSFR as a function of stellar mass and spectral type. Spectral types and stellar mass estimates come from the catalogue of Rowan-Robinson et al. (2008).}
\label{fig:oliver2009}
\end{figure}

Using the larger SWIRE dataset, in combination with photometric redshift and stellar mass estimates from Rowan-Robinson et al. (2008), Oliver et al. in prep investigate the SSFR as not only a function of mass and redshift, but also optical spectral type. Figure \ref{fig:oliver2009} shows the results of this analysis. While a similar increase of the SSFR with redshift is observed, in contrast with the other results some dependence of the evolution with stellar mass is observed, once the sample is broken down by spectral type. In particular the elliptical samples appear to show a strong decrease in SSFR with increasing mass, while late-type galaxies show either no trend or a slight increase. 

\section{Environments of star forming galaxies}
It is well accepted that environment plays a key role in the evolution of galaxies (e.g. Moore et al. 1996; Dressler et al. 1997; Farrah et al. 2004; Hogg et al. 2004; Kauffmann et al. 2004; Bamford et al. 2009.) However actually quantifying this role via observations has proven difficult. {\it Spitzer} observations over wide areas has done much to alleviate this by allowing consistent samples of galaxies across a range of environments, redshifts, and spectral types, to be assembled.

Elbaz et al. (2007) investigated the environment of star forming galaxies at $z\sim1$ in the GOODS field. Using a ``counts-in-cells'' approach the local galaxy number density was estimated by the number of galaxies within boxes of comoving size 1.5 Mpc. Combining this information with star formation rates estimated from deep MIPS 24$\mu$m observations the relationship between star formation rate and environment is presented. Curiously a strong increase in star formation rate with increasing galaxy number density is observed, in stark contrast to what is seen in the local Universe ($z\sim0.1$) by surveys such as 2dFGRS or SDSS.

Taking a similar, but philosphically different, approach Caputi et al. 2009 investigated the properties of close neighbours ($<1$ Mpc or within 500 kms$^{-1}$) to 24$\mu$m selected LIRGs and ULIRGs at $z\sim0.8$ in the COSMOS field. Via measurements of the D$_n$(4000) indices of the neighbouring galaxies they find that LIRGs lie in regions with a significantly greater fraction of passive galaxies than ULIRGs; 42\% of LIRG neighbours have D$_n$(4000)$>1.4$, compared to 24.5\% for ULIRGs. Interestingly both of these numbers are significantly different from the fraction measured for a control sample of randomly selected galaxies; $\sim 30$\%. By looking at the mass-weighted number density around their LIRG and ULIRG samples Caputi et al. find that LIRGs at $z\sim0.8$ lie in overdense regions, while ULIRGs lie in underdense regions. They speculate that this is the underlying cause of the difference in the stellar populations of LIRG and ULIRG neighbours.

To investigate this more quantitatively Frost et al. in prep use the large area of the SWIRE survey, in combination with photo-$z$'s and SED fitting from Rowan-Robinson et al. (2008), to measure the spatial correlation function for SWIRE galaxies as a function of spectral type, redshift and stellar mass.

\begin{figure}[!ht]
\begin{center}
   \resizebox{1\hsize}{!}{
     \includegraphics*{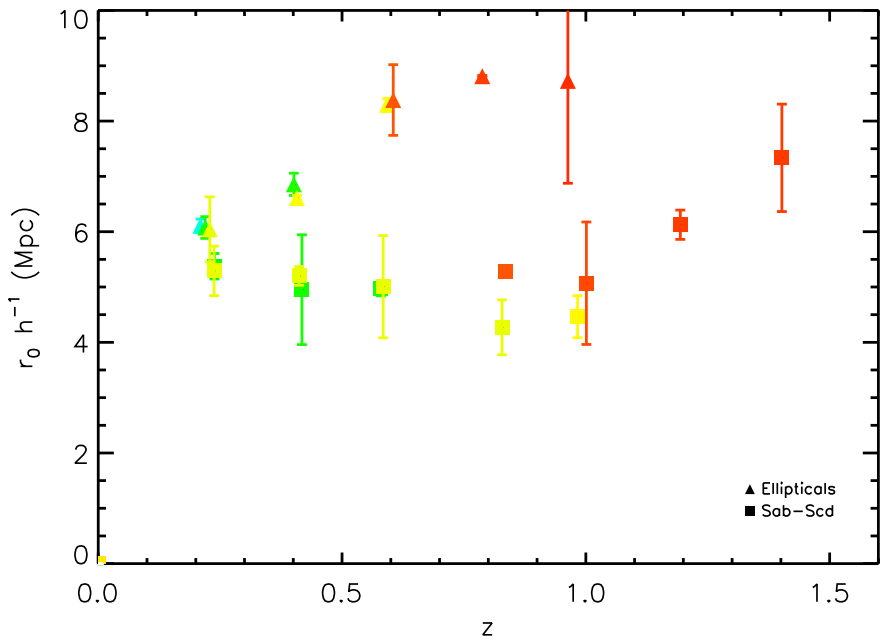}
   }
\end{center}
\caption{The comoving clustering strength $r_0$ as a function of redshift for Elliptical (triangles) and Sab-Scd galaxies (squares).  Multiple stellar mass ranges are shown for each type; 10$^{10}$M$_{\odot}<$ M$_{*}<10^{10.5}$M$_{\odot}$ (light blue), 10$^{10.5}$M$_{\odot}<$ M$_*<10^{11}$M$_{\odot}$ (green), 10$^{11}$M$_{\odot}<$ M$_*<10^{11.5}$ M$_{\odot}$ (yellow), 10$^{11.5}$M$_{\odot}<$ M$_*<10^{11.8}$M$_{\odot}$ (red).}
\label{fig:frost2009}
\end{figure}
Figure \ref{fig:frost2009} shows the clustering strength for early and late-type galaxies as a function of redshift and stellar mass. It is clear that the colour-density relation was in place at $z=1$ in agreement with \cite{coil08} and \cite{mccracken08}.  Interestingly the clustering of early and late-type galaxies does not scale with stellar mass, in agreement with \cite{coil08} but at odds with low redshift measurements \cite{norberg02} and \cite{budavari03}. This implies that at higher redshifts elliptical galaxies are found in the same environments regardless of their stellar mass content.  Frost et al also find the clustering of elliptical galaxies to increase steeply with redshift over $0.1<z<1.0$ whereas it decreases slightly for spirals over the same redshift range.  The clustering of massive star forming spirals is found to increase strongly with redshift for $z>1.0$, in qualitative agreement with the work discussed above.

\begin{figure}[!ht]
\begin{center}
   \resizebox{\hsize}{!}{
     \includegraphics*{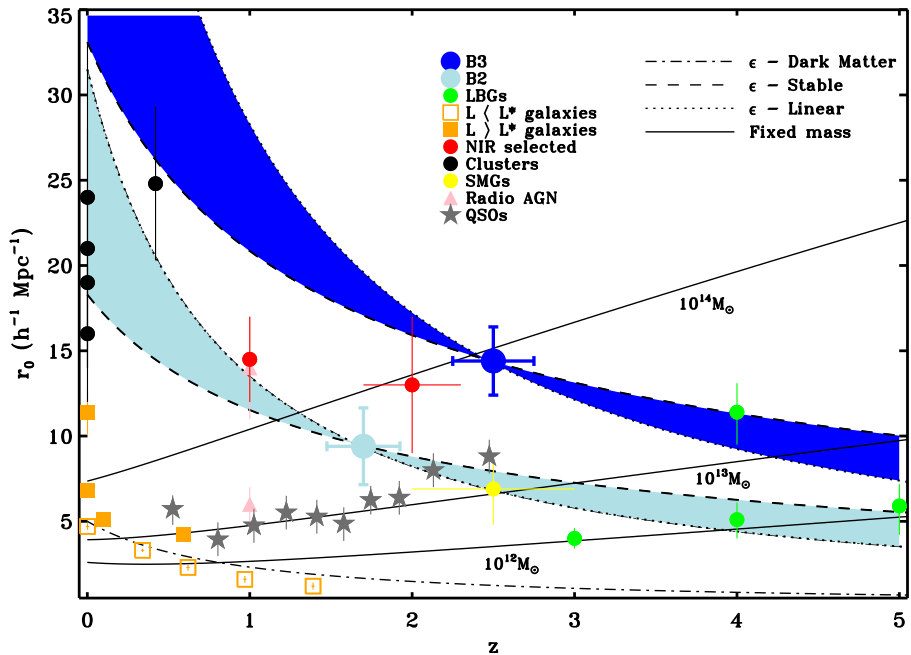}
   }
\end{center}
\caption{Comoving correlation length ($r_0$) vs. redshift for the {\it Spitzer} selected high-$z$ sources as well as a compliation of other measures (Farrah et al. 2006 and references therein). Assuming a range of dark matter haloe evolution it is clear that high-$z$ ULIRGs evolve to occupy massive ($>10^{13}$M$_{\odot}$) DM haloes.}
\label{fig:farrah2006}
\end{figure}

At high-$z$ Farrah et al. (2006) use the effect of redshifting the 1.6$\mu$m stellar ``bump'' feature through the IRAC bands to isolate samples of galaxies in the SWIRE survey at $z\sim1.7$ and $z\sim2.5$. Farrah et al. select Bump-2 \& Bump-3 sources (sources which ``peak'' in the IRAC 4.5$\mu$m and 5.8$\mu$m band, respectively) which are also found to be brighter than 400$\mu$Jy at 24$\mu$m. These selections roughly translate to a stellar mass limit of 10$^{11}$M$_{\odot}$ and IR luminosity limit of 10$^{12}$M$_{\odot}$ i.e. the most massive ULIRGs at high-$z$. Via measurement of their spatial correlation, both samples are found to have moderately strong clustering strengths, with a clustering length of $r_0=6.14\pm0.84\ h^{-1}$ Mpc for Bump-2's ($1.5<z<2.0$) and $r_0=5.36\pm1.28\ h^{-1}$ Mpc for Bump-3's ($2<z<3$). Interesting this compares well with measurements of clustering of high-$z$ QSO's, as shown in Figure \ref{fig:farrah2006}. Assuming a realistic model for dark matter halo evolution to $z=0$ shows that these galaxies will occupy high density environments ($>10^{13}$M$_{\odot}$ DM halo mass). Thus these galaxies are the likely progenitors of $>L_*$ ellipticals at $z=0$.

Similar levels of strong clustering are found for analogous samples of high-$z$ ULIRGS. Brodwin et al. (2008) measure the correlation length of Dust Obscured Galaxies (DOGs) at $1.5<z<2.5$ to be $r_0=7.4^{+1.27}_{-0.84}$. They find the clustering strength of DOGs is strongly dependent, with the most luminous sources ($F_{24}>0.6$mJy) having a correlation length of $r_0=12.97^{+4.26}_{-2.64}$. Similarly Magliocchetti et al. (2008) find a correlation length of $r_0=8.5^{+1.5}_{-1.8}$ for a photo-$z$ selected sample of $z \ge 1.6$ $F_{24}>0.4$ mJy sources. Thus it is clear that ULIRGs at $z\sim 2$ inhabit regions which are the progenitors of $z=0$ clusters and groups.

\section{Conclusion}
While the results outlined here form an impressive resume for the achievements of {\it Spitzer} they represent only a small fraction of the vast accomplishments of the observatory in this field. In particular highlights excluded from this review include important work such as that on local galaxies performed by the SINGS survey (Kennicutt et al. 2003), and the vast array of discoveries made via mid IR spectroscopic observations with the IRS instrument (e.g. Brandl et al. 2006; Smith et al. 2006; Yan et al. 2007; Spoon et al. 2008; Pope et al. 2008; Farrah et al, 2008; amongt many others). 

While much has already been learnt from the large survey programs conducted with {\it Spitzer} the true legacy nature of these datasets will soon become apparent with projects such as {\it Herschel} and SCUBA-2 set to re-observe {\it Spitzer} legacy fields. While these new datasets represent the next frontier in far-IR/sub-mm astronomy the existing {\it Spitzer} data will be crucial in interpreting these data. In this way, combined with a new generation of large {\it Spitzer} surveys undertaken as part of the ``warm'' mission, {\it Spitzer} will continue to play a key role in the study of star forming galaxies and their evolution for years to come.

\acknowledgements 
IR thanks the conference organisors for their kind invitation and funding support.\\


\end{document}